\documentstyle[aps,floats]{revtex}

\begin{document}
\input epsf
\renewcommand{\topfraction}{1.0}
\twocolumn[\hsize\textwidth\columnwidth\hsize\csname@twocolumnfalse\endcsname

\title{Detection of low energy solar neutrinos with HPGermanium}
\author{L. Baudis and H.V. Klapdor--Kleingrothaus}
\address{Max--Planck--Institut f\"ur Kernphysik, \\
P.O.Box 10 39 80, D--69029 Heidelberg, Germany}
\maketitle

\begin{abstract}
The potential of the GENIUS proposal \cite{klap} to measure the
spectrum of low energy solar neutrinos in real time is studied.
The detection reaction is elastic neutrino-electron scattering
$\nu + e \rightarrow \nu + e$.  The energy resolution for 
detecting the recoil electrons is about 0.3 \%, the energy threshold
is a few keV. The expected number of events for a
target of one ton of natural germanium is 3.6 events/day for
pp-neutrinos and 1.3 events/day for $^7$Be-neutrinos, calculated 
in the standard solar model (BP98 \cite{SSM}).
It should be feasible to achieve a background low enough to measure 
the low energy solar neutrino spectrum. 

\end{abstract}
\vskip2pc]

\newpage

All solar neutrino experiments measure a deficit of the neutrino flux 
compared to the predictions of the standard solar model (SSM) \cite{SSM}.
These predictions have recently been confirmed by helioseismology
\cite{basu} to a high 
precision, a fact which strongly disfavours astrophysical solutions proposed 
to explain the discrepancies between the theory and measurements.
An explanation of the results of solar neutrino experiments 
seems to require new physics beyond the standard model of weak interaction.
The most robust predictions of the SSM are for the pp, pep and $^7$Be 
fluxes, the pp-flux being most strongly constrained by the solar 
luminosity. In this context, a real time measurement of the pp-flux 
would be of crucial importance, since any deviation from the predicted 
flux would be a signature for neutrino flavour oscillations.
So far, there exist three proposals to measure the pp-flux in real time, 
HERON \cite{heron}, HELLAZ \cite{hellaz} and LENS \cite{lens}, 
all of them still in a stage of development.
In this letter, we explore the potential of the GENIUS project to 
measure the pp- and $^7$Be-neutrino flux by the elastic scattering process 
$\nu$ +  e$^- \rightarrow$ $\nu$ +  e$^-$.

GENIUS is a detector proposed to search for dark matter WIMPs and 
for the neutrinoless double 
beta decay using ionization in natural and enriched $^{76}$Ge HPGe
detectors, respectively \cite{klap}.
In a first  step (dark matter version), GENIUS would operate about 40
natural Ge detectors  
(100 kg) in a 12$\times$12 m tank filled with liquid nitrogen. The 
nitrogen acts both as cooling medium for the Ge crystals and as 
shielding against the natural radioactivity of the environment.
For almost complete covering of the MSSM parameter space predicted for 
neutralinos as dark matter candidates, a background counting rate of 
0.01 events/kg y keV in the energy region below 100 keV is required.
Such a low background opens the possibility to measure the pp and 
the  $^7$Be  neutrino flux in real time with the specific high energy
resolution  
of Ge detectors and  an energy threshold of a few keV.

The reaction used to detect solar neutrinos is the elastic neutrino 
electron scattering: $\nu$ +  e$^- \rightarrow$ $\nu$ +  e$^-$.
The maximum electron recoil energy is 261 keV for the pp-neutrinos and 665 
keV for the  $^7$Be-neutrinos \cite{bahc89}. The recoil electrons can
be detected  
through their ionization in a HPGe detector with an energy resolution 
of 0.3\%.
The detection rates for the pp and $^7$Be-fluxes, calculated for the 
SSM \cite{SSM}, are R$_{pp}\simeq$ 70 SNU and R$_{^7Be}\simeq$ 26 SNU (1 SNU = 
10$^{-36}$/(s target atom)). For one ton of natural Ge (corresponding 
to 6$\times$ 10$^{29}$ electrons), the total rates are R$_{pp}\simeq$ 
3.6 events/day  and R$_{^7Be}\simeq$ 1.3 events/day, assuming the
detection of all electrons. This is about ten times higher than the rates 
in present radiochemical Ga-experiments.
The event rates for full $\nu_e \rightarrow \nu_{\mu}$ conversion 
are 0.96 events/day for pp-neutrinos and 0.28 events/day for
$^7$Be-neutrinos.
GENIUS can measure only the energy distribution of the recoiling 
electrons, whereas the energy of the incoming neutrinos is not directly 
determined. However, due to the excellent energy resolution of the 
detectors and the 
difference in the elastic scattering cross section of electron and 
muon neutrinos, a comparison of the energy spectrum of recoiling 
electrons with the theoretical prediction of the SSM can be made.
Due to its relatively high counting rate, GENIUS will be able to test the MSW
flavour conversion solution via the day-night modulation of the
neutrino flux and the vacuum-oscillation solution via the seasonal
flux variation.

The possibility to operate 'naked' Ge-crystals directly in liquid 
nitrogen has been investigated in three consecutive technical studies
\cite{jochen,nim99,prep}.
It has been shown that the performance of the detectors is as good as 
for conventionally operated Ge diodes. The energy resolution of a 400 
g detector is 1 keV at 300 keV and the energy threshold is 2.5 keV. 
No microphonic events beyond the threshold, no interference between 
two or more crystals and no signal deterioration up to cable lengths 
of 6 m between crystals and FET were observed.
To estimate the expected background, detailed Monte Carlo simulations 
were performed \cite{nim99}.
Table \ref{backlist} shows the results of  simulations  
for the main simulated components together with the assumptions about the 
material radiopurities and used fluxes.
A background counting rate of 10$^{-2}$ events/kg y keV, as achieved
in \cite{nim99},
would still be a factor of two higher than the neutrino induced
signal, which is 5.8$\times$10$^{-3}$ events/kg y keV in the energy
region from 0 to 260 keV. 
Therefore, a clear neutrino signal above background, assuming
no background subtraction (with the exception of the
2$\nu\beta\beta$-decay induced signal), requires some additional
assumptions in comparison to \cite{nim99}.
First, in order to obtain a high count rate, one ton of natural Ge
($\sim$ 300 Ge detectors) has to be used. This represents a very low
target mass compared to other solar neutrino detectors. The high
counting rate is a consequence of the low energy threshold for
single electron recoil detection (11 keV in the worst case), making
GENIUS sensitive to a very large part of the pp-neutrino flux 
(about 10$^4$ times higher than the $^8$B-flux).
The tank diameter has to be increased to 13 m in order to provide
sufficient shielding from the natural radioactivity of the Gran Sasso
walls (alternatively an outer water shielding could be used). 
Regarding muon showers, the count rate is reduced by a factor
100 with respect to \cite{nim99}, due to the anticoincidence of the
300 Ge detectors \cite{jochen}.
For the intrinsic contamination of the liquid nitrogen  the
simulation was updated with the last measurements for $^{238}$U and $^{232}$Th of
the liquid scintillator of Borexino \cite{borex} and with the
$^{222}$Rn contamination measurements of liquid nitrogen \cite{rau} (see Table
\ref{backlist}).
Regarding the holder system, 130 g of holder material per detector
were assumed in \cite{nim99}. 
A new technical study revealed the possibility to use only 3 
g of material per detector in total \cite{prep}.
In the actual simulation, 13 g material per detector were 
assumed.
From the produced radionuclides by muon generated neutron interactions 
in the liquid nitrogen, only the excited $^{14}$C$^*$ nuclei yield a non-negligible
count rate \cite{nim99}. However these gamma rays can be discriminated by
the anticoincidence with a muon veto shield.
Not considered in \cite{nim99} were neutron interactions in the Ge
detectors themselves. In 1 ton of natural Germanium, 2.3$\times$10$^2$ neutrons/y due to 
muon interactions are produced. For the low energy region the 
most signifikant reaction is the $^{70}$Ge(n,$\gamma$)$^{71}$Ge capture
reaction. $^{71}$Ge decays through EC (100\%) with T$_{1/2}$ = 11.43 d
and Q$_{\rm EC}$ = 229.4 keV \cite{firestone} and can not be discriminated by the
anticoincidence method. 

\begin{figure}
\epsfxsize=15cm
\centerline{\epsffile{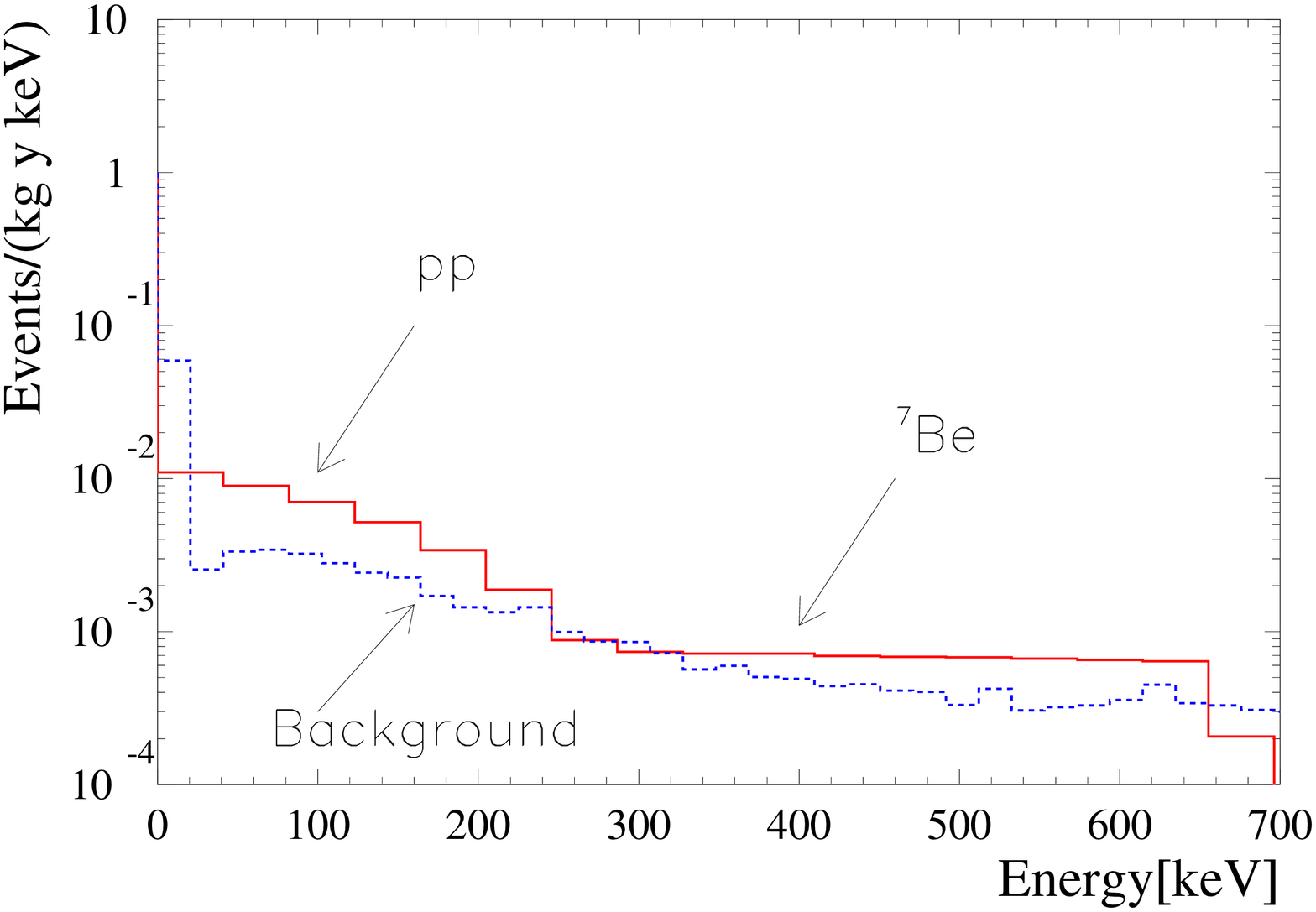}}
\caption{Simulated spectra of the low energy neutrino signal (in the SSM) and the
  total background in GENIUS (1 ton of natural germanium).}
\label{spektrum}
\end{figure}

The cosmic activation 
of the Ge crystals during their production and transportation at sea
level accounts for the still most dangerous background. 
In \cite{nim99}, 10 days of exposure at sea level and 3 years
deactivation in low level environment were assumed.
To reduce the cosmogenic  
background to an acceptable level, 
a maximum activation time of 1 day and a deactivation time of 5 years
is required. This requires
production of the detectors in underground facilities and a short
transportation time with strong shielding.
The double beta decay of $^{76}$Ge (7.8\% in natural Ge) yields 3$\times$10$^{-2}$
events/kg y keV in the 11-260 keV energy region. This dominates by far
the other background sources.
However, due to the knowledge of the half-life of the decay \cite{heimo1}, the
spectral shape and the amount of $^{76}$Ge nuclei in the detector,
this component can be calculated and subtracted.
Fig \ref{spektrum} shows the expected electron recoil spectrum
obtained by MC simulations, using the total pp- and $^7$Be-flux from
\cite{SSM} and the electron spectrum of $\nu$-e scattering from
\cite{bahc95}, together with the expected background spectrum.

With the above assumptions, the signal to background ratio is about 2:1. 
Such a ratio would be sufficient for an unambiguously detection of electron
recoils from solar pp and $^7$Be neutrinos. The good energy resolution 
of the detector and the timing information of the signals would
further help to discriminate the signal from background events, due
to the expected shape of the recoil electrons (which depends directly
on the neutrino energy spectrum) and due to a possible time variation
of the neutrino induced signal.

In summary, we investigated the capability of the GENIUS project to detect the
solar pp- and $^7$Be-neutrino flux via electron-neutrino elastic
scattering reactions. The detection rate of pp-neutrinos is 3.6
events/day and 1.3 events/day for $^7$Be neutrinos in the SSM
\cite{SSM}. 
The required
background rate for a 2:1 signal to background ratio is about
3$\times$10$^{-3}$ events/kg y keV in the energy region from 11 to 260
keV. Although this imposes
very strong purity restrictions for all the detector components, a
liquid nitrogen shielding of 13 m in diameter and production of the Germanium 
detectors below ground, it should be feasible to achieve such a low
background level.
The advantages of the experiment are the well understood detection technique
(ionization in a HPGe detector), the excellent energy resolution (1 keV 
at 300 keV), low energy threshold (about 11 keV) and the measurement
of the recoiling electrons in real time.
The pp-flux is most accurately predicted by solar models and strongly
constrained by the solar luminosity and helioseismological
measurements.
A measurement of the pp- and $^7$Be-neutrino flux by GENIUS could
provide an essential contribution to solve the solar
neutrino puzzle within a reasonable time scale.

\acknowledgments 
L.B. was supported by 
the Graduiertenkolleg of the University of Heidelberg.
She would like to thank Rocky Kolb for stimulating discussions
in Aspen, September 1998.

\onecolumn

\begin{table}
\begin{center}
\begin{tabular}{lllr}
Source & Component & Assumption & Events/(kg y keV)]     \\
       &           &            & (11-260 keV) \\
\hline
LiN, intrinsic  & $^{238}$U, $^{232}$T, $^{40}$K & 3.5, 4.4,
10$\times$10$^{-16}$g/g & 3.6$\times$10$^{-4}$\\ 
contamination & $^{222}$Rn  & 0.5 $\mu$Bq/m$^3$ &  2.5$\times$10$^{-5}$ \\        
\hline
Steel vessel   & U/Th & 10$^{-8}$g/g  &  4.5$\times$10$^{-5}$ \\
\hline
Holder system  & U/Th & 10$^{-13}$g/g; 13g material/det. &8$\times$10$^{-5}$  \\
\hline
Surrounding    & Gammas      & GS flux; tank: 13$\times$13 m &9$\times$10$^{-4}$ \\
               & Neutrons    &  GS flux & 3$\times$10$^{-4}$\\
               & Muon showers &  GS flux; muon veto &
               7.2$\times$10$^{-6}$ \\
               & $\mu$ $\rightarrow$ n ($^{71}$Ge)& 2.3$\times$10$^2$
               capt. in nat. Ge/y  &  5$\times$10$^{-4}$  \\
               \hline
Cosmogens     &$^{54}$Mn,$^{57}$Co,$^{60}$Co,$^{63}$Ni,
$^{65}$Zn,$^{68}$Ge & 1d activ., 5y deactiv.  & 8$\times10^{-4}$ \\
\hline

Total          & &  &3$\times$10$^{-3}$  \\
\hline 
\hline 
Signal (pp + $^7$Be) & & & 5.8$\times$10$^{-3}$  \\
\end{tabular}
\end{center}
\caption{Background and neutrino induced signal in GENIUS 
in the energy region from 11 keV to 260 keV.} 
\label{backlist}
\end{table}

\end{document}